\documentclass[aps, pra, twocolumn, superscriptaddress]{revtex4}
\usepackage{graphicx}
\usepackage{amsmath}
\usepackage{amssymb}
\usepackage{bm}

\usepackage[colorlinks, linkcolor=blue, urlcolor=magenta, citecolor=blue]{hyperref}
\begin{document}

\title{Observation of topology of non-Hermitian systems without chiral symmetry}

\author{Shuo Wang}
\affiliation{Guangdong Provincial Key Laboratory of Quantum Metrology and Sensing \& School of Physics and Astronomy, Sun Yat-Sen University (Zhuhai Campus), Zhuhai 519082, China.}

\author{Zhengjie Kang}
\affiliation{Guangdong Provincial Key Laboratory of Quantum Metrology and Sensing \& School of Physics and Astronomy, Sun Yat-Sen University (Zhuhai Campus), Zhuhai 519082, China.}

\author{Hao Li}
\affiliation{Guangdong Provincial Key Laboratory of Quantum Metrology and Sensing \& School of Physics and Astronomy, Sun Yat-Sen University (Zhuhai Campus), Zhuhai 519082, China.}

\author{Jiaojiao Li}
\affiliation{Guangdong Provincial Key Laboratory of Quantum Metrology and Sensing \& School of Physics and Astronomy, Sun Yat-Sen University (Zhuhai Campus), Zhuhai 519082, China.}

\author{Yuanjie Zhang}
\affiliation{Guangdong Provincial Key Laboratory of Quantum Metrology and Sensing \& School of Physics and Astronomy, Sun Yat-Sen University (Zhuhai Campus), Zhuhai 519082, China.}

\author{Zhihuang Luo}
\email{luozhih5@mail.sysu.edu.cn}
\affiliation{Guangdong Provincial Key Laboratory of Quantum Metrology and Sensing \& School of Physics and Astronomy, Sun Yat-Sen University (Zhuhai Campus), Zhuhai 519082, China.}

\begin{abstract}
Topological invariants are crucial for characterizing topological systems. However, experimentally measuring them presents a significant challenge, especially in non-Hermitian systems where the biorthogonal eigenvectors are often necessary. We propose a general approach for measuring the topological invariants of one-dimensional non-Hermitian systems, which can be derived from the spin textures of right eigenstates. By utilizing a dilation method, we realize a non-Hermitian system without chiral symmetry on a two-qubit nuclear magnetic resonance system and measure the winding number associated with the eigenstates. In addition to examining the topology of the eigenstates, our experiment also reveals the topological structure of the energy band, which differs from that in chiral systems. Our work paves the way for further exploration of complex topological properties in non-Hermitian systems without chiral symmetry.
\end{abstract}

\maketitle

\section{Introduction}

Topological phases of matter play a central role in condensed matter physics and fault-tolerant quantum computing~\cite{Hasan2010Colloquium, Qi2011Topological, Wen2017Colloquium, Kitaev2003Fault, Nayak2008Non}. Recent studies on topological phases have extended to non-Hermitian systems and attracted a great interest in both theoretical~\cite{Ghatak_2019,Chiu_2016,Zhou_2019,Kawabata_2019,Bergholtz_2021,Ding_2022,Gong_2018,Shen_2018,Zhang_2020,Yao_2018,Yi_2020,Okuma_2020,Zhang_2022,Aquino_2023,Rui_2023,Zhang_2021,He_2023,Jiang_2018,Yin_2018,Liang_2013,Wojcik_2020,Hu_2021,Lee_2016,Leykam_2017,Lieu_2018,Shunyu_2018,Yokomizo_2019,Kunst_2018,Borgnia_2020,Zhu_2020,Huang_2023} and experimental domains~\cite{wengang_2021,Wu_2023,Kai_2021,Zhang_2023,Zeuner_2015,Liang_2023,Doppler_2016,Xiao_2020,Liu_2021,Wang_2021,cao_2023,Rui_2021,Helbig_2020,Hu_2023,Lin_2022,Xu_2016,Huili_2022,Yu_2022}. Different from the Hermitian systems, non-Hermitian systems exhibit unique properties such as complex eigenvalues~\cite{Ashida_2020}, biorthonornal eigenvectors~\cite{Brody_2014}, exceptional points~\cite{Bergholtz_2021,Ding_2022,Doppler_2016,Xiao_2020,Liu_2021}, and non-Hermitian skin effects~\cite{Yao_2018,Yi_2020,Okuma_2020,Zhang_2022}. These novel phenomena provide valuable insights into non-Hermitian systems. 
Due to the global nature, topological invariants such as Chern number and winding number can be employed to characterize different topological phases~\cite{Hatsugai_1993,Thouless_1982,cui_2013,Qi_2011,asboth_2016}. However, non-Hermiticity presents challenges for the realization of non-Hermitian systems and the measurement of topological invariants based on biorthogonal eigenvectors, when the definitions of topological invariants in Hermitian systems generalize to non-Hermitian systems~\cite{Jiang_2018,Yin_2018}. The method for measuring the topological invariants in non-Hermitian systems is urgent and may be a prerequisite for exploring non-Hermitian topological phenomena in experiments.

In the non-Hermitian context, measurements of topological invariants are often associated with the right and left eigenstates, governed by \(H\) and \(H^{\dagger}\), respectively. Recent theoretical works proposed that the winding number can be measured through the spin textures in non-Hermitian systems~\cite{Zhu_2020,Huang_2023}. In these methods, both the right and left eigenstates are required, which can impose a burden on experiments, because we must separately implement \(H\) and \(H^{\dagger}\) to obtain the evolutions of the right and left eigenstates. On the other hand, it is well established that symmetry is crucial in the study of topological properties~\cite{Chiu_2016,Zhou_2019,Kawabata_2019}. For non-Hermitian systems, the topological characteristics differ significantly depending on whether or not chiral symmetry is present~\cite{Jiang_2018,Yin_2018}. However, recent experimental studies have primarily focused on measuring the topological invariants of non-Hermitian systems with chiral symmetry~\cite{wengang_2021,Wu_2023}, while experimental research on non-chiral systems is still lacking.

In this paper, we report an experimental measurement of the topological invariants in one-dimensional (1D) non-Hermitian systems without chiral symmetry. We employ a dilation method to realize the evolution of a general non-Hermitian system on a two-qubit nuclear magnetic resonance (NMR) system. 
Starting with a right state of the non-Hermitian Hamiltonian, the dilated system evolves under the Hermitian Hamiltonian. We observe the time evolutions of right-state spin textures for a non-Hermitian system through the projection measurement onto the dilated system.
We propose a method to measure the winding number of a non-Hermitian system using the right-eigenstate spin textures. The right-eigenstate spin textures and related complex eigenvalues can be extracted by fitting the observed time evolutions of right-state spin textures. Using this approach, the experiment only requires the evolution under \(H\), without the simulation of \(H^{\dagger}\). 
Furthermore, we demonstrate that when the chiral symmetry of a non-Hermitian Hamiltonian is broken, the topological invariants of its eigenstates and eigenvalues exhibit different phase transition boundaries~\cite{Jiang_2018}. This contrasts with systems possessing chiral symmetry, where the boundaries of both coincide~\cite{Yin_2018}.

\section{\label{sec:1}Theory}

The Hamiltonian of a 1D two-band non-Hermitian system can generally be expressed as
\begin{eqnarray}\label{eq: Hamiltonian}
    H(k)=\bm{h}(k)\cdot \bm{\sigma}.
\end{eqnarray}
where $\bm{h}(k) = (h_x(k), h_y(k), h_z(k))$ is a complex vector field that depends on the quasimomentum $k$, and \(\bm{\sigma} = (\sigma_x, \sigma_y, \sigma_z)\) denotes the Pauli matrices. Noting that if $h_z=0$, the Hamiltonian \eqref{eq: Hamiltonian} exhibits chiral symmetry, i.e., \(\sigma_z H(k) \sigma_z =-H(k)\). However, when $h_z\ne 0$, the Hamiltonian \eqref{eq: Hamiltonian} breaks the chiral symmetry. Due to the non-Hermiticity of the system, $H(k)\neq H^{\dag}(k)$, the distinct right and left eigenstates are defined by \(H(k)|\varphi_{\mu}^R(k)\rangle = E_{\mu}(k)|\varphi_{\mu}^R(k)\rangle\) and \(H^{\dagger}(k)|\varphi_{\mu}^L(k)\rangle = E^*_{\mu}(k)|\varphi_{\mu}^L(k)\rangle\), respectively. Here \(E_{\mu}(k)=\mu\sqrt{h_x^2(k)+h_y^2(k)+h_z^2(k)}\) are the complex eigenvalues, with $\mu = \pm$ indexing the two bands. In constract to Hermitian systems, neither \(|\varphi^R_{\mu}(k)\rangle\) nor \(|\varphi^L_{\mu}(k)\rangle\) form an orthogonal basis. Instead, they combine to create a pair of biorthogonal vectors that satisfy: \(\langle \varphi_{\mu}^L(k) | \varphi_{\mu'}^R(k) \rangle = \delta_{\mu\mu'}\) and $\sum_\mu |\varphi_\mu^R(k)\rangle\langle\varphi_\mu^L(k)| = 1$ by the normalization~\cite{Brody_2014}. The explicit forms of the right and left biorthogonal eigenstates of the Hamiltonian \eqref{eq: Hamiltonian} are given by
\begin{eqnarray}\label{eq:eigen1}
    \begin{aligned}
    |\varphi_{\mu}^R(k)\rangle &=\frac{1}{\sqrt{2E_{\mu}(E_{\mu}-h_z)}}(h_x-ih_y,\quad E_{\mu}-h_z)^T, \\
    \langle\varphi_{\mu}^L(k)| &=\frac{1}{\sqrt{2E_{\mu}(E_{\mu}-h_z)}}(h_x+ih_y,\quad E_{\mu}-h_z).
    \end{aligned}
\end{eqnarray}
If the system starts at an initial right state \(|\psi^R(k,0)\rangle= \sum_\mu c_\mu|\varphi_\mu^R(k)\rangle\) or its  related left state \(\langle\psi^L(k,0)|= \sum_\mu c^{*}_\mu\langle\varphi_\mu^L(k)|\), their time evolutions governed by $H(k)$ and $H^{\dagger}(k)$ are described by
\begin{eqnarray}\label{eq: psiRt}
    \begin{aligned}
    |\psi^R(k,t)\rangle &= \sum_{\mu} c_{\mu}(k) e^{-iE_{\mu}(k)t} |\varphi^R_{\mu}(k)\rangle, \\
    \langle\psi^L(k,t)| &= \sum_{\mu} c_{\mu}^*(k) e^{iE_{\mu}^*(k)t} \langle\varphi^L_{\mu}(k)|.
    \end{aligned}
\end{eqnarray}

To characterize the properties of non-Hermitian systems, it is essential to consider the topological invariants, such as the winding number. One can readily generalize the definition of winding number that is related to the Berry phase in Hermitian systems to non-Hermitian systems, which can be written as~\cite{Yin_2018, Lieu_2018}
\begin{eqnarray}\label{eq: windingnumber}
    w_{\mu}=\frac{1}{\pi}\oint dk\langle\varphi_\mu^L(k)|i\partial_k|\varphi_\mu^R(k)\rangle,
\end{eqnarray}
where $\mu=\pm$ indicates the band labels.
For Hermitian systems, we have $|\varphi_\mu^L(k)\rangle = |\varphi_\mu^R(k)\rangle$, whereas in non-Hermitian systems, $|\varphi_\mu^L(k)\rangle \neq |\varphi_\mu^R(k)\rangle$, making it challenging to detect the $w_\mu$ in experiments. 


By substituting the explicit forms of $|\varphi_{\mu}^R(k)\rangle$ and $\langle \varphi^L_{\mu}(k)|$ into Eq. \eqref{eq: windingnumber}, we represent $w_{\mu}$ as follows
\begin{eqnarray}
    w_{\mu}=\frac{1}{2\pi}\oint_Sdk\frac{h_x\partial_kh_y-h_y\partial_kh_x}{E_\mu(E_{\mu}-h_z)}.
\end{eqnarray}
For the case of chiral symmetry (i.e., $h_z = 0$), it can be found that the $w_\mu$ for each band takes the same value, i.e., $w_+ = w_-$. But for the general case without chiral symmetry, the $w_\mu$ is not a quantized number and $w_+ \neq w_-$, showing that the $w_{\mu}$ is no longer topological invariant. Nevertheless, their sum,
\begin{equation}\label{eq:w2}
     w_t=w_+ + w_- =\frac{1}{\pi}\oint_S\partial_k\phi_{yx}dk,
\end{equation}
has been demonstrated to be a topological invariant~\cite{Yin_2018}, where \(\phi_{yx}=\arctan(h_y/h_x)\) is a complex angle.
Since \(\mathrm{Im}(\phi_{yx})\) is a real continuous periodic function of \(k\), we have \(\oint_c\partial_k \text{Im}(\phi_{yx})dk=0\), which means that only \(\mathrm{Re}(\phi_{yx})\) contributes to the integral of Eq.(\ref{eq:w2}).

To measure the winding number $w_t$, recent theoretical work~\cite{Zhu_2020} demonstrated that the real part of \(\phi_{yx}\) can be decomposed as
\begin{equation}\label{eq: RePhiRL}
	\mathrm{Re}[\phi_{yx}(k)] = \frac{1}{2}\left[\phi_{yx}^{RR}(k) + \phi_{yx}^{LL}(k)\right] + n\frac{\pi}{2},
\end{equation}
where $n$ is an integer, and
\begin{equation}\label{eq: phiRL}
   \phi_{yx}^{\beta\beta} = \arctan\left(\frac{\overline{\langle\psi^{\beta}(k,t)|\sigma_y|\psi^{\beta}(k,t)\rangle}}{\overline{\langle\psi^{\beta}(k,t)|\sigma_x|\psi^{\beta}(k,t)\rangle}}\right).
\end{equation}
with $\beta = R, L$,  and the long-time average of spin texture is defined as
\(\overline{\langle \psi^\beta(k,t) | \sigma_\alpha | \psi^\beta(k,t) \rangle} = \lim_{T \to \infty} \frac{1}{T} \int_0^T \frac{\langle \psi^\beta(k,t) | \sigma_\alpha | \psi^\beta(k,t) \rangle}{\langle \psi^\beta(k,t) | \psi^\beta(k,t) \rangle} dt\) ($\alpha = x, y, z$). 
Since the spin textures are observable, equation (\ref{eq: RePhiRL}) implies that the winding number $w_t$ of a non-Hermitian system can be experimentally detected from $\phi_{yx}^{RR}$ and $\phi_{yx}^{LL}$, which are related to the dynamics of the right and left states governed by \(H(k)\) and \(H^{\dagger}(k)\), respectively. Consequently, it is insufficient to extract the $w_t$ from the time evolution of either $H(k)$ or $H^{\dagger}(k)$ alone. Furthermore, the realization of $H^{\dagger}(k)$ presents a practical obstacle in the experiment.

\begin{figure}
    \centering
    \includegraphics[width=0.49\textwidth]{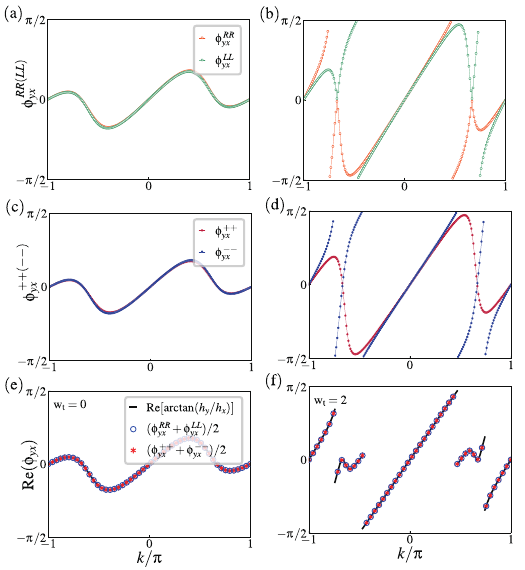} 
    \caption{(a,b) \(\phi_{yx}^{RR}\) and \(\phi_{yx}^{LL} \) as a function of \(k\), calculated from the long time average of right-right and left-left spin textures. (c,d) \(\phi_{yx}^{++}\) and \(\phi_{yx}^{--} \) as a function of \(k\), computed from the spin textures of the two right eigenstates. (e,f) \(\mathrm{Re}(\phi_{yx})\), defined by $\phi_{yx}\equiv\text{actan}(h_y/h_x)$, is equal to \((\phi_{yx}^{RR}+\phi_{yx}^{LL})/2 \) and \((\phi_{yx}^{++}+\phi_{yx}^{--})/2\). Here parameters are set to \(J_0=3, J_{1,2}=1\), \(\delta=0.3\) with $w_t = 0$ (left panel) and \(J_0=1, J_{1,2}=1\), \(\delta=0.3\) with $w_t = 2$ (right panel).}
    \label{fig: Fig1}
\end{figure}

Similarly, $\text{Re}(\phi_{yx})$ can be rewritten in another way (see the proof in Appendix \ref{Ap:A}):
\begin{equation}\label{eq: RePhiPM}
	 \text{Re}[\phi_{yx}(k)]= \frac{1}{2}\left[\phi_{yx}^{++}(k) + \phi_{yx}^{--}(k)\right] + n\frac{\pi}{2},
\end{equation}
where
\begin{equation}\label{eq: phiPM}
	 \phi_{yx}^{\mu\mu}(k) = \arctan\left(\frac{{\langle\varphi^R_\mu(k)|\sigma_y|\varphi^R_\mu(k)\rangle}}{{\langle\varphi^R_\mu(k)|\sigma_x|\varphi^R_\mu(k)\rangle}}\right), 
\end{equation}
with $\mu=\pm$. Figure~\ref{fig: Fig1} shows the different results of Eqs. \eqref{eq: RePhiRL}-\eqref{eq: phiPM}, where we take an example with \(h_x=J_0+J_1\cos(k)+J_2\cos(2k)\), \(h_y=J_1\sin(k)+J_2\sin(2k)-i\delta\) and \(h_z=0.5\). The parameters are set to \(J_0=3, J_{1,2}=1\), \(\delta=0.3\) for Figs. \ref{fig: Fig1}(a), \ref{fig: Fig1}(c), and \ref{fig: Fig1}(e) with $w_t = 0$, and \(J_{0,1,2}=1\), \(\delta=0.3\) for Figs. \ref{fig: Fig1}(b), \ref{fig: Fig1}(d), and \ref{fig: Fig1}(f) with $w_t = 2$. Though \(\phi_{yx}^{++}\) and \(\phi_{yx}^{--}\) in Fig. \ref{fig: Fig1}(d) are different from \(\phi_{yx}^{RR}\) and \(\phi_{yx}^{LL}\) in Fig. \ref{fig: Fig1}(b), the averages of them are same with the theoretical value of $\text{Re}[\phi_{yx}]$, as shown in Fig. \ref{fig: Fig1}(f).
Based on the above results, $\mathrm{Re}[\phi_{yx}(k)]$ can be obtained solely from the right-right spin textures \(\langle\varphi^R_\pm(k)|\sigma_\alpha|\varphi^R_\pm(k)\rangle\) ($\alpha = x, y$), whose dynamics are determined by only \(H(k)\). This simplifies the process of measuring the winding number $w_t$ of a general non-Hermitian Hamiltonian \eqref{eq: Hamiltonian} in experiments, as there is no longer a need to implement the \(H^{\dagger}(k)\). Both \(\langle\varphi^R_\pm(k)|\sigma_\alpha|\varphi^R_\pm(k)\rangle\) for $\alpha=x \text{ or } y$ can be obtained simultaneously by fitting the time evolution of spin textures $\langle\psi^R(k, t)|\sigma_\alpha|\psi^R(k, t)\rangle$, if $c_\pm\neq 0$ (see the results in Sec. \ref{sec: results}). 

\section{\label{sec: experiment}Experiment}

\begin{figure}
    \centering
    \includegraphics[width=0.5\textwidth]{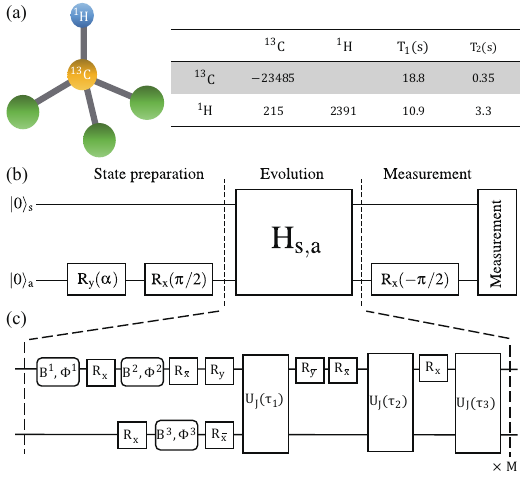} 
    \caption{(a) Molecular structure of  \(^{13}C\)-labeled chloroform consisting of a $^{13}$C and $^{1}$H nuclear spins, and its Hamiltonian parameters. The diagonal and off-diagonal elements represent the chemical shifts and J-coupling constants (in Hz). The $T_1$ and $T_2$ are spin-lattice relaxation times and spin-spin relaxation times (in seconds). (b) Quantum circuit for realizing the dynamical evolution of a general non-Hermitian system using the dilation method. The $^{13}$C and $^1$H are used as the system qubit and ancillary qubit, respectively. $R_y(\alpha)$ denotes the single-qubit rotation $\alpha$ along the $y$ direction, with \(\alpha= 2\arctan\eta_0\). (c) Pulse sequence for realizing the dilated Hamiltonian $H_{s, a}$. $B^i$ and $\Phi^i$ ($i=1, 2, 3$) given by Appendix. \ref{sec: dilatedMethod} represent the amplitudes and phases of the control Hamiltonian \eqref{eq: RF}, $U_J(\tau_j) = \text{exp}[-i\frac{\pi J}{2}\sigma_z^1\sigma_z^2\tau_j]$ is the free evolution of the NMR Hamiltonian \eqref{eq: NMR}, where we set $\nu_i^{\text{RF}}= \nu_i$ in our experiment, and $\tau_j=2\gamma_j(t_m)\tau/\pi J$.}
    \label{fig: Fig2}
\end{figure}

The experiment is performed on a Bruker AVANCE 300\(\mathrm{MHz}\) spectrometer with a 2-qubit sample \(^{13}C\)-labeled chloroform dissolved in acetone-d6, whose molecular structure and Hamiltonian parameters are illustrated in Fig. \ref{fig: Fig2}(a). Its NMR Hamiltonian in a rotating frame is 
 \begin{equation}\label{eq: NMR}
 H_{\mathrm{NMR}}=-\sum_i^2\pi(\nu_i-\nu_i^{\text{RF}})\sigma_z^i + \frac{\pi J}{2}\sigma_{z}^{1}\sigma_{z}^{2}.  
 \end{equation}
 Here $\nu_i$ and $\nu_i^{\text{RF}}$ ($i=1,2$) represent the chemical shifts and the reference frequencies of RF fields, and the $J$ is the coupling strength between $^{13}$C and $^{1}$H.
Each nuclear spin can be individually controlled by radio-frequency (RF) pulses. The corresponding control Hamiltonian reads
  \begin{equation}\label{eq: RF}
 H_{\mathrm{c}}= \sum_{i=1}^{2}\pi B_{i}(\cos\Phi_{i}\sigma_{x}^{i}+\sin\Phi_{i}\sigma_{y}^{i}),    
 \end{equation}
 where \(B_{i}\) and \(\Phi_i\) are amplitudes and phases of the RF fields. By tuning the control parameters $B_i$ and $\Phi_i$ of RF fields, we can simulate the dynamical evolution of a general two-qubit system.
 
  \begin{figure*}
    \centering
    \includegraphics[width=0.95\textwidth]{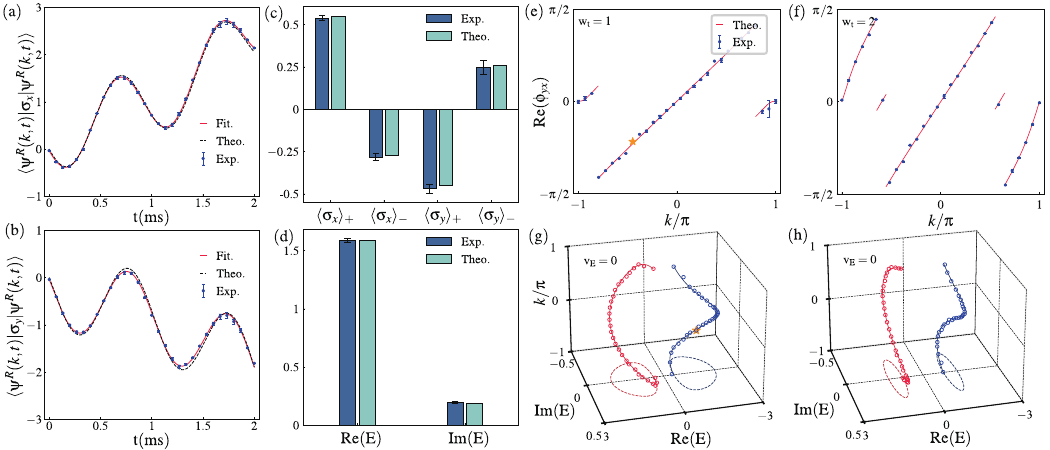} 
    \caption{(a-b) Time evolutions of spin textures \(\langle\psi^R(k,t)|\sigma_\alpha|\psi^R(k,t)\rangle\) for $\alpha = x, y$ and $k=-0.448\pi$, corresponding to the points labeled by orange stars in (e) and (g). (c) eigenstate spin textures \(\langle\sigma_\alpha\rangle_\pm = \langle \varphi^R_\pm(k)|\sigma_\alpha|\varphi^R_\pm(k)\) for $\alpha = x \text{ and } y$, obtained by fitting the curves of (a) and (b), respectively. (d) Real and imaginary parts of the complex eigenvalue \(E\), resulted from the same fitting procedure. (e-f) \(\text{Re}(\phi)_{yx}\) as a function of \(k\) with \(w_t=1\) and \(w_t=2\). (g-h) $\text{Re}(E)$ and $\text{Im}(E)$ as a function of $k$ with \(\nu_E=0\). Parameters are chosen as $J_0 = 1, J_1 =1, \delta = 0.3$ for (e) and (g), and $J_0 = 0.3, J_1 =1, \delta = 0.3$ for (f) and (h), respectively.}
    \label{fig: Fig3}
\end{figure*}
 
 We consider a non-Hermitian Hamiltonian \eqref{eq: Hamiltonian} with a complex vector field: 
 \begin{equation}\label{eq: vectorField}
  h_x=J_0+J_1\cos(k), h_y=J_1 \sin(k)-i\delta, h_z=0.5.
 \end{equation}
 It is a nonchiral system because of $h_z\neq 0$. The phase diagram exhibits two different topological phases with $w_t=1$ for \(J_0 =1, J_{1}=1, \delta=0.3\), and $w_t = 2$ when \(J_0 =0.3, J_{1}=1, \delta=0.3\).
To implement the non-Hermitian Hamiltonian in the experiment, we exploit a dilation method~\cite{Wu_2019,Liu_2021}. The dilated Hermitian Hamiltonian obtained from the non-Hermitian Hamiltonian \eqref{eq: Hamiltonian} can be expressed as  
\begin{equation}\label{eq: Hsa}
   H_{s,a}(t)=\Lambda(t)\otimes I+\Gamma(t) \otimes \sigma_z,
\end{equation}
where $\Lambda(t) = \sum_{\alpha=0}^3\lambda_\alpha(t)\sigma_\alpha$ and $\Gamma(t) = \sum_{\alpha=0}^3\gamma_\alpha(t)\sigma_\alpha$, and $\sigma_0$ denotes the identity operator $I$. The coefficients of $\lambda_\alpha$ and $\gamma_\alpha$ can be found in Appendix. \ref{sec: dilatedMethod}.
By introducing an ancillary qubit, the dilated state evolves as
\begin{equation}\label{eq: Psisa}
    |{\Psi}(k, t)\rangle_{s,a} =|\psi^R(k, t)\rangle_s |-\rangle_a +\eta(t)|\psi^R(k, t)\rangle_s |+\rangle_a,
\end{equation} 
under the \(H_{s,a}(t)\), where $|\psi^{R}(k, t)\rangle_s$ is the right state of the non-Hermitian system (see Eq. \eqref{eq: psiRt}), \(|\pm \rangle_a\)  are eigenstates of \(\sigma_y\) of the ancillary qubit and \(\eta(t)\) is an appropriate linear operator~\cite{Wu_2019,Liu_2021}. 

Figure \ref{fig: Fig2}(a) shows the quantum circuit for realizing the dynamical evolution of the non-Hermitian Hamiltonian using the dilation method. Starting with the initial state: \(|0\rangle_s|0\rangle_a\), the system is prepared to the dilated state: $|{\Psi}(k, 0)\rangle_{s,a} =|\psi^R(k, 0)\rangle_s |-\rangle_a +\eta(0)|\psi^R(k, 0)\rangle_s |+\rangle_a$ by two single-qubit rotations $R_y(\alpha)$ (\(\alpha= 2\arctan\eta_0\)) along the $y$ direction and $R_x(\pi/2)$ along the $x$ direction acting the ancillary qubit,  and evolves under the dilated Hermitian Hamiltonian $H_{s, a}$, which can be implemented by pulse sequence of Fig. \ref{fig: Fig2}(c). After a $R_x(-\pi/2)$, the final state of the dilated system is 
\begin{equation}\label{eq: Psisaf}
    |{\Psi}(k, t)\rangle_{s,a}^f =|\psi^R(k, t)\rangle_s |0\rangle_a +\eta(t)|\psi^R(k, t)\rangle_s |1\rangle_a.
\end{equation} 
Thus, we can obtain the time evolution of the right state \(|\psi^R_s(t)\rangle\) by projecting onto the \(|0\rangle_a\langle 0|\) subspace.
In experiments, we employ the Gradient Ascent Pulse Engineering (GRAPE) technique~\cite{KHANEJA_2005} to optimize the quantum circuit in Fig. \ref{fig: Fig2}(b) with high precision. 
The observable operators $\sigma_\alpha\otimes|0\rangle_a\langle0|$ with $\alpha=x,y$ are chosen for the measurements of $\langle\psi^R(k, t)|\sigma_\alpha|\psi^R(k, t)\rangle=\sum_{\mu,\mu'}^\pm c_\mu^* c_{\mu'} e^{i (E_\mu^* - E_{\mu'}) t} \langle \varphi^R_\mu|\sigma_\alpha|\varphi^R_{\mu'}\rangle$. If the initial state $|\psi^R(k, 0)\rangle = \sum_\mu^\pm c_\mu |\varphi_\mu^R\rangle$ is prepared to satisfy that $c_\pm \neq 0$, $ \langle \varphi^R_\pm|\sigma_\alpha|\varphi^R_\pm\rangle$ for $\alpha = x$ or $y$ can be extracted by fitting the time evolution of spin texture $\langle\psi^R(k, t)|\sigma_\alpha|\psi^R(k, t)\rangle$. We also note that the real and imaginary parts of the complex eigenvalue $\text{Re}(E)$ and $\text{Im}(E)$ can be obtained from the same fitting procedure, because of $E_+ = -E_-$. 
 
 \section{\label{sec: results} Results}

Figures \ref{fig: Fig3}(a) and \ref{fig: Fig3}(b) show the time evolutions of spin textures \(\langle\psi^R(k,t)|\sigma_\alpha|\psi^R(k,t)\rangle\) for $\alpha = x, y$ and $k=-0.448\pi$. Here, the non-Hermitian Hamiltonian parameters were taken as $J_0 = 1, J_1 =1, \delta = 0.3$, corresponding to the winding number $w_t = 1$. The experimental results are in agreement with the theoretical predictions, confirming that the dilated method for simulating the dynamical evolution of the non-Hermitian Hamiltonian \eqref{eq: Hamiltonian} is feasible in the experiment. From the fitting results of Figs. \ref{fig: Fig3}(a) and \ref{fig: Fig3}(b), the eigenstate spin textures \(\langle\sigma_\alpha\rangle_\pm = \langle \varphi^R_\pm(k)|\sigma_\alpha|\varphi^R_\pm(k)\) for $\alpha = x, y$ were extracted, as presented in Fig. \ref{fig: Fig3}(c). We now get a measured data point of $\text{Re}[\phi_{yx}(k)]$ for $k = -0.448\pi$, according to Eqs. \eqref{eq: RePhiPM} and \eqref{eq: phiPM}. Also, we obtained the real and imaginary parts of the eigenvalues depicted in Fig. \ref{fig: Fig3}(d), from the fitting results.
 
Using the same method above, we measured \(\text{Re}[\phi_{yx}(k)]\) for $k\in [-\pi, \pi]$. The results of 
two distinct topological phases with $w_t =1$ and $w_t=2$ are shown in Figs. \ref{fig: Fig3}(e) and \ref{fig: Fig3}(f), respectively.
 From these results, it is evident that we successfully measured the topological invariants \(w_t\) of a non-Hermitian Hamiltonian that breaks chiral symmetry, without the need of the left states whose dynamical evolutions are governed by \(H^{\dagger}\).

In addition to the nontrivial topology of the eigenstates, the topological structures of the energy bands were also observed. The eigenvalues of a non-Hermitian Hamiltonian are generally complex, and its band structures also exhibit topological properties, which can be described by another winding number \(v_E\)~\cite{Gong_2018,Shen_2018},
\begin{eqnarray}\label{eq: ve}
    \nu_E=\sum_{\mu}^\pm\int_{-\pi}^\pi\frac{dk}{2\pi}\partial_k\arg E_\mu(k).
\end{eqnarray}
where \(\arg E_\mu(k)\) is the argument of the complex energy \(E_\mu(k)=\mu\sqrt{1+J_0^2-\delta^2+h_z^2+2J_0\cos k-i2\delta\sin k}\). 
Figures \ref{fig: Fig3} (g) and \ref{fig: Fig3}(h) show the band structures of our model with both $\nu_E =0$, as the projections of the energy bands on the complex energy plane form two separate circles.
It is worth noting that there are no band-touching points, implying the absence of a phase transition, which is different from the results with nonzero winding numbers \(w_t\)s corresponding to two distinct topological phases. 

In contrast to chiral systems, the phase boundaries characterized by \(w_{t}\) and \(\nu_E \) coincide~\cite{Jiang_2018}. However, in the absence of the chiral symmetry, the phase boundaries of \(v_E\) correspond to the band-touching points of the non-chiral system, but no band touching occurs at the phase boundaries of \(w_t\). This is quite different from the chiral non-Hermitian Hamiltonian. Our experiments exhibit the unusual phenomenon of a non-chiral system by measuring both eigenstate topological invariant \(w_t\) and eigenvalue topological invariant \(v_E\).  The error analysis can be found in Appendix. \ref{sec: errorAnalysis}.

\section{\label{sec:4} Conclusion}

We presented a general scheme for measuring the winding numbers of 1D non-Hermitian systems. Unlike previous works~\cite{Zhu_2020, Huang_2023}, our approach does not rely on left eigenstates governed by \(H^{\dagger}\), offering an experimental advantage. By employing the dilated method~\cite{Wu_2019,Wu_2023}, we realized a non-Hermitian system without chiral symmetry on a two-qubit NMR system and measured the winding numbers using our scheme, where the right-eigenstate spin textures of two bands were simultaneously extracted by fitting the observed time evolutions of spin textures. Meanwhile, we also obtained the complex eigenvalues for two distinct topological phases from the same fitting procedure. 
Our experimental results demonstrated a discrepancy between the phase boundaries of phase diagrams characterized by the two independent topological invariants \(w_t\) and \(\nu_E\), due to the breaking of chiral symmetry. In future work, extending our scheme to explore the topology of higher-dimensional non-Hermitian models, such as measuring Chern numbers in two-dimensional systems, would be particularly interesting but also experimentally challenging, because of the extensive spin textures to be measured in high dimensions.

\section*{Acknowledgements}
This work was supported by the National Natural Science Foundation (Grant No. 11805008), Guangdong Basic and Applied Basic Research Foundation (Grant No. 2024A1515011406), Fundamental Research Funds for the Central Universities, Sun Yat-Sen University (Grant No. 23qnpy63), Guangdong Provincial Key Laboratory (Grant No. 2019B121203005).

\appendix

\section{Relations between azimuthal angle and eigenstate spin textures}\label{Ap:A}

According to Eq.(\ref{eq:eigen1}),  the right eigenstates of Hamiltonian (\ref{eq: Hamiltonian}) can also be rewritten as:
\begin{equation}\label{eq: A1}
    \begin{aligned}
    |\varphi_{+}^R\rangle &=(e^{-i\phi_{yx}}\cos\frac{\beta}{2}, \sin\frac{\beta}{2})^T, \\ 
    |\varphi_{-}^R\rangle &=(e^{-i\phi_{yx}}\sin\frac{\beta}{2}, -\cos\frac{\beta}{2})^T,
    \end{aligned}
\end{equation}
where
\begin{equation}\label{eq: A2}
    \cos \beta= \frac{h_z}{\sqrt{h_x^2+h_y^2+h_z^2}}, \quad  e^{i\phi_{yx}}=\frac{h_x+ih_y}{\sqrt{h_x^2+h_y^2}}.
\end{equation}
Here \(\phi_{yx}=\arctan(h_y/h_x)\) is the azimuthal angle discussed in the main text. We first define the quantity $\phi_{yx}^{\mu\mu}$ ($\mu=\pm$) that satisfy
\begin{equation}\label{eq: A3}
    \tan\phi^{\mu\mu}_{yx} \equiv \frac{\langle \varphi_\mu^R |\sigma_y|\varphi_\mu^R \rangle}{\langle \varphi^R_\mu|\sigma_x|\varphi^R_\mu \rangle}.
    \end{equation}
 By substituting Eqs. \eqref{eq: A1} and \eqref{eq: A2} into Eq. \eqref{eq: A3}, we can obtain
\begin{equation}
\begin{aligned}\label{eq:spt}
    \tan \phi_{yx}^{++}&=i\frac{e^{-i\phi_{yx}}\mathcal{S}^*-e^{i\phi^*_{yx}}\mathcal{S}}{e^{-i\phi_{yx}}\mathcal{S}^*+e^{i\phi^*_{yx}}\mathcal{S}}, \\
    \tan \phi_{yx}^{--}&=i\frac{e^{-i\phi_{yx}}\mathcal{S}-e^{i\phi^*_{yx}}\mathcal{S}^*}{e^{-i\phi_{yx}}\mathcal{S}+e^{i\phi^*_{yx}}\mathcal{S}^*}.  
\end{aligned}
\end{equation}
with $\mathcal{S} = \sin \frac{\beta}{2} \cos \frac{\beta^{*}}{2}$. Therefore, we have
\begin{equation}
\begin{aligned}
    \tan(\phi^{++}_{yx}+\phi^{--}_{yx}) & =\frac{\tan\phi^{++}_{yx}+\tan\phi^{--}_{yx}}{1-\tan \phi^{++}_{yx}\tan\phi^{--}_{yx}} \\
    &= i\frac{e^{-i2\phi_{yx}}-e^{i2\phi^*_{yx}}}
    {e^{-i2\phi_{yx}}+e^{i2\phi^*_{yx}}}\\ &= \tan[2\mathrm{Re}(\phi_{yx})],
\end{aligned}
\end{equation}
and it is easy to derive the relationship of Eq. \eqref{eq: phiPM} in the main text, i.e.,
\begin{equation}\label{eq:phyx}
    \mathrm{Re}(\phi_{yx})=\frac{1}{2}(\phi^{++}_{yx}+\phi^{--}_{yx})+ \frac{n\pi}{2}.
\end{equation}
Note that only the right eigenstates have been used in the above derivation. 

 \section{The dilated method for realizing a non-Hermitian system}\label{sec: dilatedMethod}
 
\begin{figure}
    \centering
    \includegraphics[width=0.49\textwidth]{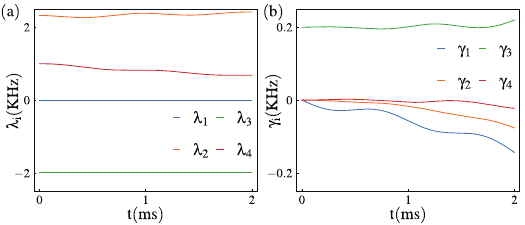} 
    \caption{Time-dependent control parameters $\lambda_\alpha(k, t)$ (a) and $\gamma_\alpha(k, t)$ (b) of the dilated Hermitian Hamiltonian \eqref{eq: Hsa2}, where $\alpha = 0, 1, 2, 3$, and \(k=-0.448\pi\).}
    \label{fig: lambdaGamma}
\end{figure}

 To implement a non-Hermitian Hamiltonian in the experiment, we exploit the dilation method~\cite{Wu_2019,Liu_2021}. The Hermitian Hamiltonian dilated by an ancillary qubit can be expressed as  
\begin{equation}
\label{eq:Hsa}
   H_{s,a}(t)=\Lambda(t)\otimes I+\Gamma(t) \otimes \sigma_z,
\end{equation}
with
\begin{equation}
\begin{aligned}
\Lambda(t)&=\{H_s(t)+[i\frac{d}{dt}\eta(t)+\eta(t)H_s(t)]\eta(t)\}M^{-1}(t),\\
\Gamma(t) &= i \left[ H_{s}(t) \eta(t) - \eta(t) H_{s}(t) - i \frac{d}{dt} \eta(t) \right] M^{-1}(t).
\end{aligned} 
\end{equation}
Here \(H_s(t)\) is the Hamitonian of non-Hermitian system, \(M(t) = \mathcal{T} e^{-i \int_{0}^{t} H_s^{\dagger}(t) dt} M(0) \bar{\mathcal{T}} e^{i \int_{0}^{t} H_s(t) dt}\), and \(\eta(t) = \sqrt{M(t) - I}\), where \(\mathcal{T}\) is time-ordering operator (\(\hbar\) is set to 1 for simplicity), \(M(0) = (\eta_0^2 + 1) I\),  \(\eta_0\) needs to be set appropriately to keep \( M(t) - I \) positive for all \( t \), and \( I \) is identity operator. Considering a general 1D two-band non-Hermitian Hamiltonian \eqref{eq: Hamiltonian},
the dilated Hamiltonian can be expanded in terms of Pauli operators:
\begin{equation}\label{eq: Hsa2}
    H_{s,a}(k, t) = \sum_{\alpha=0}^3\lambda_\alpha(k, t)\sigma_\alpha^s \otimes I^a +\sum_{\alpha=0}^3\gamma_\alpha(k, t) \sigma_\alpha^s \otimes \sigma_z^a,
\end{equation}
where $\lambda_\alpha(k, t)$ and $ \gamma_\alpha(k, t)(\alpha = 0,1,2,3)$ are real time-dependent control parameters. For the case of $k=-0.448\pi$, the $\lambda_\alpha(k, t)$ and $\gamma_\alpha(k, t)$ for $\alpha = 0,1,2,3$ are illustrated in Fig. \ref{fig: lambdaGamma}.

We now employ the Trotter approximation formula to realize $H_{s,a}(k, t)$. The total evolution time $T$ is divided into $M$ time slices with a duration of $\tau=T/M$. The evolution for $t_m = m\tau$ can be approximated to
\begin{equation}
\begin{aligned}
        e^{-i\tau H_{s,a}(t_m)} \approx & e^{-i\tau [\lambda_2(t_m)\sigma_x^s + \lambda_3(t_m)\sigma_y^s]}e^{-i\tau [\lambda_4(t_m)\sigma_z^s+\gamma_1(t_m)\sigma_z^a]} \\
        &e^{-i\tau \gamma_2(t_m)\sigma_x^s \sigma_z^a}e^{-i\tau \gamma_3(t_m)\sigma_y^s\sigma_z^a}e^{-i\tau \gamma_4(t_m)\sigma_z^s\sigma_z^a}.
\end{aligned}
\end{equation}
By adjusting the amplitudes and phases of the RF fields (see Eq. \eqref{eq: RF}), i.e.,
\begin{equation}
\begin{aligned}
	B_s^1 &= \frac{\tau}{\pi} \sqrt{\lambda^2_2(t_m)+\lambda^2_3(t_m)}, \\
	 \Phi_s^1 &= \text{arccos}\left [\lambda_2(t_m)/\sqrt{\lambda^2_2(t_m)+\lambda^2_3(t_m)}\right ], 
\end{aligned}
\end{equation}
we can realize the evolution $e^{-i\tau [\lambda_2(t_m)\sigma_x^s + \lambda_3(t_m)\sigma_y^s]}$. 
Similarly, we set
\begin{equation}
\begin{aligned}
	B_s^2 &= \lambda_4(t_m)/\pi, \quad \Phi_s^2 = \pi/2, \\
	B_a^3 &= \gamma_1(t_m)/\pi, \quad \Phi_a^3 = \pi/2,
\end{aligned}
\end{equation}
to realize the evolutions 
\begin{equation}
\begin{aligned}
	e^{-i\tau \lambda_4(t_m)\sigma_z^s}&= R_x^s e^{-i\pi B_s^2(\cos\Phi_s^2\sigma_{x}^{s}+\sin\Phi_s^2\sigma_{y}^{s})\tau}R^s_{\bar{x}}, \\
		e^{-i\tau \gamma_1(t_m)\sigma_z^a}&= R_x^a e^{-i\pi B_s^3(\cos\Phi_s^3\sigma_{x}^{a}+\sin\Phi_s^2\sigma_{y}^{a})\tau}R^a_{\bar{x}},
\end{aligned}
\end{equation}
where $R_{x/\bar{x}}$ denotes the $\pi/2$ or $-\pi/2$ rotation along the $x$ direction. Under the free evolution of the NMR Hamiltonian \eqref{eq: NMR}, where $\nu_i^{\text{RF}} = \nu_i$ ($i=1, 2$), we have
\begin{equation}
    \begin{aligned}
        e^{-i\tau \gamma_2(t_m)\sigma_x^s \sigma_z^a} &= R_y^se^{-i\frac{\pi J}{2}\sigma_z^s\sigma^a_z \tau_1}R_{\bar{y}}^s, \\
        e^{-i\tau \gamma_3(t_m)\sigma_y^s \sigma_z^a} &= R_x^se^{-i\frac{\pi J}{2}\sigma_z^s\sigma^a_z \tau_2}R_{\bar{x}}^s, \\
        e^{-i\tau \gamma_4(t_m)\sigma_z^s \sigma_z^a} &= e^{-i\frac{\pi J}{2}\sigma_z^s\sigma^a_z \tau_3},
    \end{aligned}
\end{equation}
with the evolution times $\tau_i=2\gamma_{i+1}(t_m)\tau/\pi J$ ($i=1,2,3$). So far, we have achieved the realization of $H_{s, a}$, as shown in the pulse sequence of Fig. \ref{fig: Fig2}(c).

\section{Experiment spectra and data}\label{sec: experimentalSpectra}

\begin{figure}
    \centering
    \includegraphics[width=0.48\textwidth]{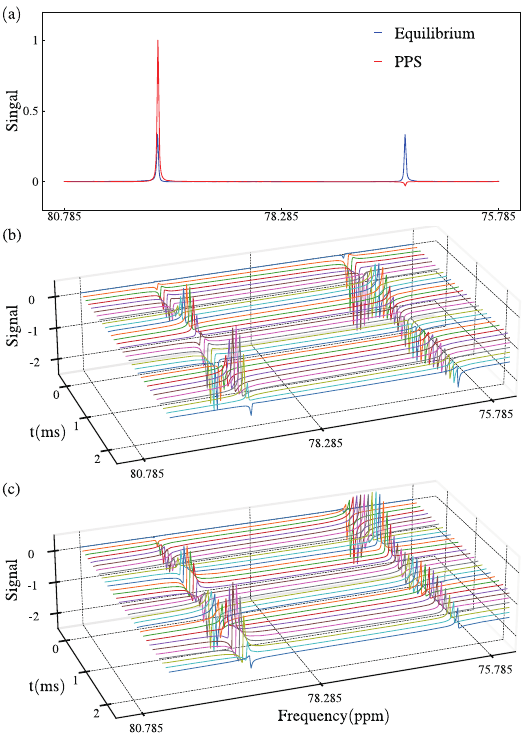} 
    \caption{(a) Experimental $^{13}$C spectra of thermal equilibrium state $\rho_{\text{eq}}$ (blue) and pseudo-pure state $\rho_{00}^{\text{PPS}}$ (red). (b-c) Experimental spectra for observing the time evolutions of spin textures \(\langle\psi^R(k,t)|\sigma_\alpha|\psi^R(k,t)\rangle\), where $\alpha = x, y$, respectively. Here $k=-0.448\pi$, and parameters are chosen as \(J_0=1\), \(J_1=1\) and \(\delta=0.3\).}
    \label{fig: NMRspectra}
\end{figure}

 At room temperature, the thermal equilibrium state of the NMR sample is 
 \begin{equation}
     \rho_{\text{eq}}\approx \frac{1-\epsilon}{4}I+\epsilon(\frac{1}{4}I+\gamma_C\sigma_z^1+\gamma_H\sigma_z^2),
 \end{equation}
where \(\epsilon\approx 10^{-5}\) is the polarization, $\gamma_C$ and $\gamma_H$ are the gyromagnetic ratio of $^{13}$C and $^1$H nuclei. We first initialize the system to the pseudo-pure state (PPS),
 \begin{equation}
     \rho_{00}^{\text{PPS}}\approx \frac{1-\epsilon}{4}I+\epsilon|00\rangle\langle00|.
 \end{equation}
 In the experiment, the $\rho_{00}^{\text{PPS}}$ was prepared from the thermal state \(\rho_{\text{eq}}\) using the line-selective method~\cite{Luo2018Experimentally, Luo2022Observation}. The experimental $^{13}$C spectra of $\rho_{\text{eq}}$ and $\rho_{00}^{\text{PPS}}$ are shown in Fig. \ref{fig: NMRspectra}(a). 
 
 \begin{figure}
    \centering
    \includegraphics[width=0.48\textwidth]{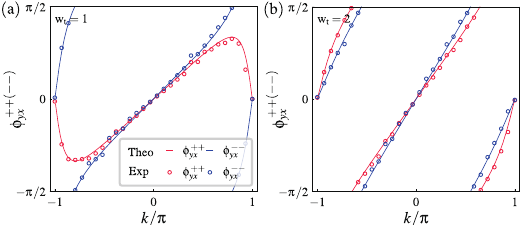} 
    \caption{The resulted \(\phi_{yx}^{++}(k)\) and \(\phi_{yx}^{--}(k) \) for two distinct topological phases with $w_t = 1$ and $w_t = 2$, by fitting the time evolutions of spin textures $\langle\psi^R(k, t)|\sigma_\alpha|\psi^R(k, t)$, where $\alpha = x, y$. The solid lines and dots are the theoretical and experimental values. }
    \label{fig: phiPMData}
\end{figure}

To observe the time evolutions of spin textures $\langle\psi^R(k, t)|\sigma_\alpha|\psi^R(k, t)\rangle$ ($\alpha = x, y$), we performed the measurements of $\sigma^s_\alpha\otimes|0\rangle_a\langle0|$ on the dilated states \eqref{eq: Psisaf}. In the case of $k = -0.448\pi$, the corresponding spectra of $\langle\psi^R(k, t)|\sigma_\alpha|\psi^R(k, t)\rangle$ for different $t$ and $\alpha = x, y$ are ploted in Figs. \ref{fig: NMRspectra}(b) and \ref{fig: NMRspectra}(c), respectively. By fitting $\langle\psi^R(k, t)|\sigma_\alpha|\psi^R(k, t)\rangle$, we simultaneously obtained the $\langle\varphi^R_\pm|\sigma_\alpha|\varphi^R_\pm\rangle$ for $\alpha = x$ or $y$, and also the $\phi_{yx}^{\mu\mu}(k) = \text{arctan}[\langle\varphi_\mu^R|\sigma_y|\varphi_\mu^R\rangle/\langle\varphi_\mu^R|\sigma_x|\varphi_\mu^R\rangle]$ ($\mu = \pm$). Figures \ref{fig: phiPMData}(a) and \ref{fig: phiPMData}(b) show the results of $\phi_{yx}^{\mu\mu}(k)$ for two different topological phases with $w_t = 1$ and $w_t = 2$, respectively. Thus, we got $\text{Re}[\phi_{yx}(k)]$, as shown in Figs \ref{fig: Fig3}(e) and \ref{fig: Fig3}(f).

\section{Error analysis}\label{sec: errorAnalysis}

We calculated the root-mean-square $ \sigma^{\text{Exp/Sim}} = \sqrt{\Sigma_i^N(\langle \sigma_{\alpha} \rangle^{\text{Exp/Sim}}_i - \langle \sigma_{\alpha} \rangle^{\text{Th}}_i )^2/N}$ between the measured or simulated spin textures and the theoretical values, where $\alpha=x,y$. The final state given by the simulation is $\rho_f^{\text{sim}} = U_{\text{Evo}}G_zU_{\text{PPS}}\rho_{\text{eq}}U_{\text{PPS}}^{\dagger}G_z U_{\text{Evo}}^{\dagger}$, where $G_z$ is a gradient magnetic field that can eliminate the off-diagonal elements of the density matrix. The spin textures come from the simulation are $\langle \sigma_\alpha \rangle =tr[\rho_f^{\text{sim}}(\sigma_\alpha\otimes I +\sigma_\alpha \otimes \sigma_z)]/2$ ($\alpha = x, y$).  When considering the influence of a specific step, we simulate the $1\%$ noise of pulses, while the rest is implemented through theoretical unitary operations. The results are listed in Table. \ref{tab:1}, where $\sigma_{\text{Sim}}^{\text{Tot}}$ represents the root mean square error of all pulses, and $\sigma_{\text{Read}}=\sigma_{\text{Exp}}-\sigma_{\text{Sim}}^{\text{Tot}}$ evaluates the readout error which came from spectral integrals. From Table. \ref{tab:1}, we find that the main errors mainly come from the preparation of the initial state, dynamical evolution, and readout. 

\begin{table}\label{tab:1}
\begin{ruledtabular}
\begin{tabular}{cccccc}
  & $\sigma_{\text{Exp}}$ & $\sigma_{\text{Sim}}^{\text{Tot}}$ & $\sigma_{\text{Sim}}^{\text{PPS}}$ & $\sigma_{\text{Sim}}^{\text{Evol}}$ & $\sigma_{\text{Read}}$ \\
\colrule
 $\langle \sigma_{x} \rangle$ &       0.0391 & 0.0202 & 0.0194 & 0.0197 & 0.0189 \\
  $\langle \sigma_{y} \rangle$ &       0.0571 & 0.0219 & 0.0079 & 0.0147 & 0.0352 \\
\end{tabular}
\end{ruledtabular}
\end{table}


 \bibliographystyle{naturemag}


\end{document}